\documentstyle[twoside,fleqn,espcrc2]{article}

\def\a{\alpha}
\def\b{\beta}
\def\c{\chi}
\def\d{\delta}
\def\e{\epsilon}

\def\g{\gamma}
\def\h{\eta}

\def\j{\psi}
\def\k{\kappa}
\def\l{\lambda}
\def\m{\mu}
\def\n{\nu}
\def\o{\omega}
\def\p{\pi}
\def\q{\theta}

\def\F{\Phi}
\def\G{\Gamma}

\def\P{\Pi}

\def\S{\Sigma}

\def\frac#1#2{{\textstyle{#1\over\vphantom2\smash{\raise.20ex
        \hbox{$\scriptstyle{#2}$}}}}}           
\def\ha{\frac12}                                

\def\pl#1#2#3{Phys.~Lett.~{\bf {#1}B} (19{#2}) #3}
\def\np#1#2#3{Nucl.~Phys.~{\bf B{#1}} (19{#2}) #3}
\def\prl#1#2#3{Phys.~Rev.~Lett.~{\bf #1} (19{#2}) #3}
\def\pr#1#2#3{Phys.~Rev.~{\bf D{#1}} (19{#2}) #3}
\def\cqg#1#2#3{Class.~and Quantum Grav.~{\bf {#1}} (19{#2}) #3}

\def\mpl#1#2#3{Mod.~Phys.~Lett.~{\bf A{#1}} (19{#2}) #3}

\def\sfo{S\!F\!O}
\def\pco{P\!C\!O}
\def\ico{I\!C\!O}

\newcommand{\AmS}{{\protect\the\textfont2
  A\kern-.1667em\lower.5ex\hbox{M}\kern-.125emS}}

\title{	Integration Measure and Spectral Flow
	in the Critical $N{=}2$ String }

\author{Olaf Lechtenfeld\address{
	Institut f\"ur Theoretische Physik, Universit\"at Hannover\\
	\ \ Appelstra\ss{}e 2, D--30167 Hannover, Germany}
        \thanks{Supported by the `Deutsche Forschungsgemeinschaft'}
	}
\begin{document}

\begin{abstract}
I present the moduli space of the $(2{+}2)$-dimensional critical
closed fermionic string with two world-sheet supersymmetries.
The integration of fermionic and Maxwell moduli in the presence
of punctures yields the string measure for $n$-point amplitudes
at arbitrary genus and instanton number.
Generalized picture-changing and spectral-flow operators emerge,
connecting different instanton sectors.
Tree and loop amplitudes are computed.
\end{abstract}

\maketitle

\section{$N{=}2$ Strings}
I am reporting on new results obtained in collaboration with
Jan Bischoff, a PhD student, and Sergei Ketov, postdoc at Hannover.
Since I have to be brief today, let me refer to our preprints
for more details~\cite{klp,bkl,wen94,kl,usc95}.
When I talk about $N{=}2$ strings, I mean fermionic strings
with two world-sheet supersymmetries. Those strings are known to
be critical in flat complex spacetime ${\bf C}^{1,1}$. In this talk
I shall consider only untwisted, oriented, closed strings in flat space
and will outline their covariant path-integral quantization \`a la BRST
in the (old-fashioned) NSR description.
Most concepts and techniques should appear familiar to string folks
from ten years ago.

Strings with extended supersymmetry were invented in 1976~\cite{ade1,ade2}.
It took a while to realize that their critical flat spacetime is
{\it complex\/} two-dimensional~\cite{ft,ali}.
The spectral flow automorphism of the $N{=}2$ superconformal algebra was
discovered in 1987~\cite{schw}.
$N{=}2$ strings have only a {\it finite\/} number of massless physical
excitations, but satisfy duality nevertheless by killing all amplitudes
with more than three legs.
This curious fact was more or less established by Ooguri and Vafa~\cite{ov}.
A first partial BRST analysis of the spectrum was performed in
1992~\cite{bien}.
Subsequently, Siegel made some interesting conjectures about connections
to the $N{=}4$ string and the possibility of spacetime supersymmetry in these
theories~\cite{sie}.
Lately, there has been an increase of activity in the field
(see, e.g., \cite{lp}), triggered by the embedding of bosonic and $N{=}1$
strings into $N{=}2$ strings~\cite{bv1,ff,op,bop} and by a new formulation
using a topologically twisted $N{=}4$ string~\cite{bv2,ber}.
For older reviews on the subject, consult refs.~\cite{ma,ket}.

Let me begin by listing the world-sheet degrees of freedom and their
symmetries.
The starting point is the $N{=}2$ world-sheet supergravity action~\cite{bs}.
The extended supergravity multiplet involves the real metric $h_{\a\b}$, a
{\it complex\/} gravitino $\c_\a$ and a real Maxwell field (graviphoton)
$A_\a$,
with $\a=0,1$.
The conformal matter consists of two complex string coordinates $X^\m$ and
two complex spin $1/2$ (NSR) fermions $\j^\m$, with $\m=0,1$.

The supergravity action is invariant under $N{=}2$ super-diffeomorphisms
and $N{=}2$ super-Weyl transformations, but shows also a global target space
symmetry generated by rigid translations and a complex `Lorentz group' of
$U(1,1)\times{\bf Z}_2$~\cite{ov}.
Superconformal gauge-fixing leads to the appropriate set of chiral ghosts,
namely a real pair of anticommuting reparametrization ghosts ($b,c$),
a complex pair of commuting supersymmetry ghosts ($\b,\g$),
and a real pair of anticommuting Maxwell ghosts ($\tilde{b},\tilde{c}$),
plus their antichiral partners.
The residual $N{=}2$ superconformal symmetry is generated by the currents
$(T, G^\pm, J)$.
As usual, world-sheet fermions may be periodic or antiperiodic around the
cylinder. Due to their complex nature, however, the spin fields mapping
Neveu-Schwarz to Ramond states carry {\it double\/} spinor indices, thus
qualifying as spacetime bosons. Hence, there is no obvious room for spacetime
supersymmetry.

\section{$N{=}2$ Moduli}
$N{=}2$ super Riemann surfaces are topologically classified by two invariants,
namely the Euler number, $\chi=2{-}2g={1\over2\pi}\int R$
where $g\in{\bf Z}_+$ is the genus and $R$ denotes the curvature two-form,
as well as the instanton number or monopole charge,
$c={1\over2\pi}\int F\in{\bf Z}$, with $F$ being the field-strength two-form.
Since the total conformal and $U(1)$ anomalies disappear for $d_{\bf C}{=}2$,
the local symmetries may be used to eliminate all but finitely many
$N{=}2$ supergravity degrees of freedom for any given pair $(g,c)$.
What remains is the integration over $N{=}2$ supermoduli space, parametrized
$m=(m_h, m_\c, m_A)$ by metric, fermionic, and Maxwell moduli.

Any metric may be decomposed into a background part carrying the Euler number,
two parts given by a diffeomorphism and a Weyl transformation,
plus a moduli component, $h(m_h)$.
In complex count, the metric moduli space is spanned by $3g{-}3$ holomorphic
quadratic differentials~$h_\ell$, which are dual to the Beltrami differentials.
The latter may be generated by quasi-conformal vector fields having
jump discontinuities along (a maximal set of) closed, non-intersecting,
minimal geodesics~$\G_\ell$.

The fermionic moduli come in two varieties, since they carry a positive or
negative unit of Maxwell charge. A basis is provided by $2g{-}2{+}c$ positive
and $2g{-}2{-}c$ negative holomorphic $\frac32$-differentials,
$\c^+_j$ and $\c^-_k$, respectively.
It is convenient to choose these differentials delta-function localized in
points $z^+_j$ and $z^-_k$ on the Riemann surface.

Finally, any Maxwell field may be Hodge-decomposed into a background
carrying the instanton number, an exact and a co-exact piece, plus a
harmonic component~$A(m_A)$ representing the Maxwell moduli.
The latter lives in the complex $g$-dimensional space of
holomorphic one-forms (or flat connections).
A popular real basis consists of real one-forms $\{\a_i,\b_i\}$
dual to the standard homology basis $\{a_i,b_i\}$.
The Maxwell moduli space is then parametrized by the twists
$\{\exp i\oint_{a_i}\!A(m_A)\,,\,\exp i\oint_{b_i}\!A(m_A)\}$
around the homology cycles, showing that it is just a (real) $2g$-torus.

\section{The String Measure}
Amplitudes for $N{=}2$ closed string scattering have now been reduced to
\begin{equation}
\langle\ldots\rangle\ =\ \sum_{g=0}^\infty\k^{2-2g}\sum_{c=2-2g}^{2g-2}\!\l^c
\int\!dm\;\langle\ldots\rangle^{(m)}_{g,c}
\end{equation}
where $\k$ and $\l$ are the gravitational and Maxwell string couplings,
and the fixed-moduli correlator is given by
\begin{equation}
\langle\ldots\rangle^{(m)}_{g,c} = \int\! D[X\j bc\b\g\tilde b\tilde c]\,
{\rm AZI} \ldots e^{iS[X-\tilde c;m]} \; .
\end{equation}
Here, $S$ stands for the gauge-fixed action,
and proper superconformal gauge fixing has led to the
antighost zero-mode insertions
\begin{eqnarray}
{\rm AZI} &=& \Bigl| \prod_\ell\langle h_\ell,b\rangle \Bigr|^2 \,
\prod_i \langle\a_i,\tilde b\rangle \langle\b_i,\tilde b\rangle \\
&& \times \Bigl| \prod_j \d\bigl(\langle\c^+_j,\b^-\rangle\bigr)
\prod_k \d\bigl(\langle\c^-_k,\b^+\rangle\bigr) \Bigr|^2 \nonumber
\end{eqnarray}
using the notation $\langle\q,\o\rangle = \int\q\wedge*\o$.

The moduli $m=(m_h,m_\c,m_A)$ appear in the action
(apart from the implicit dependence in the Hodge star $*$) only linearly via
\begin{equation}
S(m)\ \sim\ \langle h,T\rangle+\langle\c^\pm,G^\mp\rangle+\langle A,J\rangle
\quad.
\end{equation}
Thus, the integrals over fermionic and Maxwell moduli can be performed!
The former produces a factor of
\begin{equation}
\Bigl| \prod_j\langle\c^+_j,G^-\rangle \prod_k\langle\c^-_k,G^+\rangle \Bigr|^2
\end{equation}
which combines with part of the AZI to
\begin{equation}
\Bigl| \prod_j \pco^-(z^+_j) \prod_k \pco^+(z^-_k) \Bigr|^2
\end{equation}
where I defined the picture-changing operators
\begin{equation}
\pco^\pm(z)\ :=\ \d\bigl(\b^\pm(z)\bigr)\, G^\pm(z)
\end{equation}
for $\c^\pm_*\sim\d(z{-}z^\pm_*)$.
These operators are well-known and needed.
Their position dependence is BRST-trivial, so they may be moved around at will.
As for the Maxwell moduli, integrating them out yields
\begin{equation}
\prod_i \d\bigl(\langle\a_i,J\rangle\bigr)\,\d\bigl(\langle\b_i,J\rangle\bigr)
\end{equation}
which joins another part of the AZI to form
\begin{equation}
\prod_i \pco^0(a_i)\,\pco^0(b_i) \; ,
\end{equation}
introducing yet another kind of picture-changing operator,
\begin{equation}
\pco^0(\g)\ :=\ \d\Bigl(\smallint_\g*J\Bigr)\;\smallint_\g*\tilde b \; ,
\end{equation}
associated with a path~$\g$.
Note that it are Kronecker $\d$s which appear in Eqs.~(8) and~(10).
$\pco^0(\g)$ depends only on the homotopy of~$\g$.
The novel insertions of~$\pco^0$ serve to project the sum over
the complete set of states flowing through a handle onto the
charge-neutral sector, as factorization from pinching the handle
demands.

All taken together, Eq.~(1) simplifies to
\begin{equation}
\langle\ldots\rangle = \sum_{g,c}\!\k^{2-2g}\l^c\!\!\int\!\!dm_h\!\!\!
\int\!\!\!D[X-\tilde c]\,{\cal M}\ldots e^{iS_0(m_h)}
\end{equation}
where the action
\begin{equation}
S_0(m_h)\ =\ S[X-\tilde c;\, m_h, m_\c{=}0, m_A{=}0]
\end{equation}
still depends on $g$ and $c$ through the classical background metric and
Maxwell field, and the string measure takes the form
\begin{eqnarray}
{\cal M} &=& \Bigl| \prod_\ell\smallint_{\G_\ell}b \Bigr|^2 \,
\prod_i \pco^0(a_i)\,\pco^0(b_i) \nonumber\\
&&\times \Bigl| \prod_j\pco^-(z_j^+) \prod_k\pco^+(z_k^-) \Bigr|^2
\end{eqnarray}
for $g\ge2$ and $|c|\le|2g{-}2|$ (sphere and torus need minor modifications).

\section{Adding Punctures}
The proper way to deal with string amplitudes is to consider the
vertex operator insertions ($\ldots$) as $N{=}2$ punctures and
integrate ``standard vertex operators'' over the moduli space of
{\it punctured\/} (super) Riemann surfaces.
The term ``standard'' refers to the canonical representation of vertex
operators, which for $N{=}2$ closed strings means total ghost number
$u=0$ and picture numbers $\p^\pm=-1$.
Since the Euler number of a $n$-punctured surface is $2{-}2g{-}n$,
each vertex operator carries a factor of~$\k^{-1}$.
The presence of $n$ external legs then has the following consequences.

Each puncture introduces an additional (complex) metric modulus, namely the
position~$z_p$ of the puncture.
The associated $b$-ghost zero-mode insertion $|\oint_p b|^2$ may be employed
to strip the $c$~ghosts off the standard vertex operator
$V^c=c\bar cW^{(1,1)}(z_p)$.
On the sphere or torus, conformal isometries yield three or one $c$ zero modes,
each of which cancels one $b$ zero mode and one $z_p$ integration,
leaving three or one unintegrated $c$-type vertices, respectively.
All other vertex locations are to be integrated over, in accordance with
the complex dimension $3g{-}3{+}n$ of extended metric moduli space.

Likewise, each puncture adds one positive and one negative (complex) fermionic
modulus, increasing the complex dimensions of the fermionic moduli spaces
to~$2g{-}2{\pm}c{+}n$.
Repeating the computation of the previous section, this simply yields
\begin{equation}
\Bigl| \prod_p \pco^-(z_p) \, \pco^+(z_p) \Bigr|^2
\end{equation}
which may be used to convert all vertex operators to the $\p^\pm{=}0$ picture.
A notable exception is again the sphere, where two Killing spinors eliminate
the fermionic moduli of two punctures.

Finally, the number of Maxwell moduli increases.
The Maxwell field is allowed to develop single poles at the punctures,
whose residues $\oint_p A$ and $\oint_p*A$ represent additional moduli.
Their complex number is $n{-}1$ since the meromorphy of $A$ forces the
sum of the residues to vanish.
The extra homology cycles dual to these $A$~twists are the contours~$c_p$
around the vertex locations~$z_p$, paired with curves connecting the punctures
to a common base point~$z_0$.
Since a contour encircling {\it all\/} vertex locations is homotopically
trivial, and since one may select one vertex location as base point,
only $n{-}1$ such pairs of curves are independent, in agreement with
the counting of moduli. Like for the string measure~${\cal M}$,
integrating out the Maxwell puncture moduli generates $2n{-}2$ insertions
of $\pco^0$ which serve to strip the $\tilde c$~ghosts off the standard vertex
$V^{\tilde c}=\tilde c\bar{\tilde c}W^{(0,0)}(z_p)$.
This leaves a single vertex operator dressed with $\tilde c\bar{\tilde c}$,
exactly what is needed to neutralize the single $\tilde c$ and $\bar{\tilde c}$
zero modes.

In total, the moduli associated with vertex punctures can be taken into account
by choosing an appropriate set of picture- and ghost-number representatives
for the vertex operators. Correlators fail to vanish only if their total ghost
number adds up to zero and their picture numbers sum to $2g{-}2{\mp}c$.

\section{Spectral Flow}
A constant shift $m_A\to m_A{+}\d m_A$ in the integration over the Maxwell
moduli should not effect any string amplitude.
On the other hand, it does contribute an additional factor of
$\exp\{{i\over\p}\langle\d A,J\rangle\}$ to the measure,
with $\d A{\equiv}A(\d m_A)$.
Using $d*J{=}0$ and the fact that $\d A$ is harmonic outside the punctures,
one computes
\begin{eqnarray}
\langle\d A,J\rangle &=& \sum_{i=1}^g \Bigl(
\oint_{a_i}\!\d A \oint_{b_i}\!*J - \oint_{b_i}\!\d A \oint_{a_i}\!*J \Bigr)
\nonumber\\ && +\ \sum_{p=1}^n \oint_{c_p}\!\d A \int_{z_0}^{z_p}\!\!*J \; .
\end{eqnarray}
Clearly, a residue ${\rm res}_p \d A=-\q$ at $z_p{=}z$ induces a twist
by the ``spectral flow operator''
\begin{equation}
\sfo(\q,z)\ :=\ \exp\Bigl\{2\,\q\int_{z_0}^z\!*J\Bigr\}
\end{equation}
at the puncture.
The harmonicity of $\d A$ implies $\sum_p{\rm res}_p \d A=0$, restricting
the puncture twists to $\sum_p\q_p=0$.
Their effect may be subsumed by twisting the vertex operators
\begin{equation}
V_p(z_p) \longrightarrow V^{(\q_p)}_p(z_p) = \sfo(\q_p,z_p)\cdot V_p(z_p)\;.
\end{equation}

A number of interesting remarks are to be made here.
First, $\sfo$ is a {\it local\/} operator, since bosonization
$*J=\frac12d\F$ yields~\cite{bkl,kl}
\begin{equation}
\sfo(\q,z)\ =\ \l^{-\q} \exp\bigl\{\q\,\F(z)\bigr\} \; ,
\end{equation}
identifying the Maxwell string coupling
\begin{equation}
\l\ =\ \exp\bigl\{\F(z_0)\bigr\} \; .
\end{equation}
Second, the dependence on the base point~$z_0$ (and on $\l$) drops out in
correlators exactly when the total twist vanishes as required,
$\sum_p\q_p{=}0$.
Third, $\sfo$ is BRST invariant but not trivially so; its {\it derivative\/},
however, is BRST trivial, so I may move $\sfo$ around at no cost.
Fourth, as a consequence, amplitudes are twist-invariant,
\begin{eqnarray}
\langle V_1^{(\q_1)}\ldots V_n^{(\q_n)}\rangle &=&
\langle V_1^{(0)}\ldots V_n^{(0)}\,\P_p\sfo(\q_p)\rangle \nonumber\\
&=& \langle V_1^{(0)}\ldots V_n^{(0)}\,\sfo(\S_p\q_p)\rangle \nonumber\\
&=& \langle V_1 \ldots V_n \rangle \; ,
\end{eqnarray}
as long as the total twist vanishes.
This is nothing but Maxwell modular invariance.
Fifth, $\sfo(\q{=}{\pm}\ha)$ turns NS into R vertices and vice versa,
explicitly establishing their physical equivalence.
Sixth, $\sfo$ seems to ruin global $U(1,1)$ invariance
since $\exp\{\q\F\}$ carries Lorentz charge proportional to~$\q$.
However, from Eqs.~(16-19) one concludes that $\l$ must transform
compensatingly under $U(1,1)$, since $\sfo$ is BRST-equivalent to
the unit operator.
Thus, the Maxwell string coupling is {\it not\/} a Lorentz scalar!
Seventh, $\sfo(m{\in}{\bf Z},z)$ creates or destroys a singular instanton
by way of a singular gauge transformation, $A\to A-m\,d\ln z$.
More precisely,
\begin{equation}
\sfo(m,z{=}0)\,|c{=}0\rangle\ =\ \l^{-m}|c{=}m\rangle \;,\quad m{\in}{\bf Z}
\end{equation}
maps from the trivial to the $m$-instanton vacuum,
with the $c$-value coming from the delta-peaked Maxwell field strength
integrated over a disc centered at~$z{=}0$.
Hence, the amplitude contributions with nonzero values of~$c$
arise from imposing a total twist of $\sum_p \q_p =c$.
Eighth, the instanton-number-changing operators
\begin{eqnarray}
\ico^\pm(z) &:=& \l^{\pm1}\,\sfo(\q{=}{\pm1},z) \\
&=& \j^{\pm,0}\j^{\pm,1}\,\d(\g^\pm)\,\d(\b^\pm)\,(1\mp2\tilde bc)(z)\nonumber
\end{eqnarray}
may be moved to the punctures,
where they change the Lorentz charges and the picture assignments
$(\p_+,\p_-)\to(\p_+{\pm}1,\p_-{\mp}1)$ of the vertex operators.
In this way, spectral flow may be used to convert any $c{=}m$ correlator
of $U(1,1)$ singlets to a $c{=}0$ correlator of {\it Lorentz-charged\/}
vertex operators.

To sum up,
I have demonstrated how Maxwell moduli transformations correspond to
the spectral flow of the $N{=}2$ superconformal algebra, and how
the latter is implemented on correlators of vertex operators.
This connection becomes even more transparent after bosonizing the
spinor matter and ghost fields~\cite{bkl,bl}.

\section{Amplitudes}
To illustrate the arguments presented above, let me compute the leading
contributions to the three- and four-point functions of the single massless
scalar particle, restricting myself to one chiral half.
Like for the bosonic string, the massless chiral three-point correlator
must be antisymmetric in the external momenta~$k_a$, $a{=}1,2,3$.
{}From two on-shell momenta $k_a^{\pm,\m}$ and $k_b^{\pm,\n}$ one can form
three antisymmetric $O(1,1)$ invariants, namely
\begin{eqnarray}
c^0_{ab} &=& \h_{\m\n}\,k_a^{[+,\m} k_b^{-],\n} \nonumber\\
d^0_{ab} &=& \e_{\m\n}\,k_a^{\{+,\m} k_b^{-\},\n} \\
c^\pm_{ab} &=& \e_{\m\n}\,k_a^{\pm,\m} k_b^{\pm,\n} \quad,\nonumber
\end{eqnarray}
of which only the first one is $U(1,1)$ invariant.

For the simplest case of $(g,c,n)=(0,0,3)$ the total picture must be
$(\p_+,\p_-)=(-2,-2)$, so a convenient choice of representatives yields
\begin{equation}
A_3^{(0,0)}(k_1,k_2,k_3)
\end{equation}
\vskip-.2in
\begin{eqnarray}
&=&\langle V(k_1,z_1)\,V(k_2,z_2)\,V(k_3,z_3)\rangle_{(0,0)}\nonumber\\
&=&\langle\tilde ccW_{(-1,-1)}(1)\,cW_{(-1,0)}(2)\,cW_{(0,-1)}(3)
\rangle_{(0,0)}\nonumber\\
&=& c^0_{23} \quad.\nonumber
\end{eqnarray}

The $c{=}1$ contribution may be calculated from twisting, say,
the third vertex operator in Eq.~(24) by $\q{=}1$.
The same result obtains when, alternatively, I use {\it untwisted\/}
vertex operators in the $(\p_+,\p_-)=(-3,-1)$ background:
\begin{equation}
A_3^{(0,1)}(k_1,k_2,k_3)
\end{equation}
\vskip-.2in
\begin{eqnarray}
&=&\langle V(k_1,z_1)\,V(k_2,z_2)\,V(k_3,z_3)\rangle_{(0,1)}\nonumber\\
&=&\langle\tilde ccW_{(-1,-1)}(1)\,cW_{(-1,0)}(2)\,cW_{(-1,0)}(3)
\rangle_{(0,1)}\nonumber\\
&=& c^+_{23} \quad.\nonumber
\end{eqnarray}
Since $|c|\le1$ for the tree-level three-point function,
I find a total of
\begin{equation}
A_3^{g=0}\ =\ \l^{-1}\,c^-_{23} + c^0_{23} + \l\,c^+_{23} \quad .
\end{equation}
Taking into account the Lorentz properties of~$\l$ it appears that
a singlet is formed from two $SU(2)$ triplets.
This $SU(2)$ being part of the Lorents group~$U(1,1)$,
I conclude that $\ico^\pm$ play the role of spacetime Lorentz generators!

It is well-known that the absence of any massive resonances
enforces the vanishing of the four-point function~\cite{ov}.
Nevertheless, it is instructive to verify this claim.
The $N{=}2$ Veneziano amplitude receives from $c{=}0$ the term
\begin{equation}
A_4^{(0,0)}(k_1,k_2,k_3,k_4)
\end{equation}
\vskip-.2in
\begin{eqnarray}
&=& \langle\tilde ccW_{(-1,0)}\,\smallint W_{(0,-1)}\,
cW_{(-1,0)}\,cW_{(0,-1)}\rangle_{(0,0)}\nonumber\\
&\propto& k_1^+{\cdot}k_2^-\,k_3^+{\cdot}k_4^-\,t\ +\
k_1^+{\cdot}k_4^-\,k_2^-{\cdot}k_3^+\,s \nonumber \\ &\propto&
c^0_{12} c^0_{34} t\ +\ c^0_{14} c^0_{32} s\ -\ 16stu \nonumber\\
&=& 0 \quad , \nonumber
\end{eqnarray}
due to the peculiar massless kinematics in ${\bf C}^{1,1}$.
The application of spectral flow extends this vanishing
to all nonzero values of~$c$.

The machinery to evaluate loop corrections is now in place.
As an example, the one-loop three-point correlators turn out as
\begin{equation}
A_3^{(1,c)}\ \propto\ \bigl(c^0_{23}\bigr)^{3-|c|}\,\bigl(c^\e_{23}\bigr)^{|c|}
\end{equation}
with $\e{=}{\rm sign}(c){=}\pm$ and $c={-}3,\ldots,+3$.
Since all amplitudes with more than three legs should vanish for any value
of $g$ and~$c$, only the one-, two- and three-point functions for $g\ge2$
are yet to compute.

The expressions for $A_3^{(g,c)}$ are to be interpreted as string loop and
instanton corrections to the effective spacetime action, which happens to
be self-dual Yang-Mills and self-dual gravity for open and closed critical
$N{=}2$ strings, respectively.

\vglue.2in\noindent
I would like to thank my coworkers, Jan Bischoff and Sergei Ketov,
and gratefully acknowledge fruitful discussions with
George Thompson and Andrei Losev.

\vfill\eject

\end{document}